%% file: main.tex
\documentclass[sigconf]{acmart}

\acmYear{2020}

\setcopyright{none}
\copyrightyear{}
\acmConference[to appear in CCSW '20]{2020 Cloud Computing Security Workshop}{November 9, 2020}{Virtual Event, USA}
\acmBooktitle{2020 Cloud Computing Security Workshop (CCSW '20), November 9, 2020, Virtual Event, USA}
\acmPrice{}
\acmDOI{}
\acmISBN{}

\author{Ralf Ramsauer}
\affiliation{University of Applied Sciences Regensburg}
\email{ralf.ramsauer@oth-regensburg.de}

\author{Lukas Bulwahn}
\affiliation{BMW AG}
\email{lukas.bulwahn@bmw.de}

\author{Daniel Lohmann}
\affiliation{University of Hanover}
\email{lohmann@sra.uni-hannover.de}

\author{Wolfgang Mauerer}
\affiliation{University of Applied Sciences Regensburg,\\Siemens Corporate Research}
\email{wolfgang.mauerer@oth-regensburg.de}

\input{includes}

\begin{document}
\input{title}

\begin{abstract}
\input{abstract}
\end{abstract}




\maketitle

\input{content}

\bibliographystyle{ACM-Reference-Format}
\bibliography{bibliography}
\clearpage
\end{document}

%% file: includes.tex
\usepackage{amsmath, amsfonts, amsthm}
\usepackage{interval}

\usepackage{balance}
\usepackage{booktabs}
\usepackage{endnotes}
\usepackage{epigraph}
\usepackage{subfig}
\usepackage{tcolorbox}
\usepackage{float}
\usepackage{xcolor}
\usepackage{calc}
\usepackage{ifthen}
\input{tikzincludes}
\usepackage{MnSymbol}
\usepackage{paralist}

%% file: title.tex
\title[The Sound of Silence]{The Sound of Silence: Mining Security Vulnerabilities from Secret Integration Channels in Open-Source Projects}

%% file: abstract.tex
Public development processes are a key characteristic of open source projects.
However, fixes for vulnerabilities are usually discussed privately among a small group of trusted maintainers, and integrated without prior public involvement.
This is supposed to prevent early disclosure, and cope with embargo and non-disclosure agreement (NDA) rules.
While regular development activities leave publicly available traces, fixes for vulnerabilities that bypass the standard process do not.

We present a data-mining based approach to detect code fragments that arise from such infringements of the standard process.
By systematically mapping public development artefacts to source code repositories, we can exclude regular process activities, and infer irregularities that stem from non-public integration channels.
For the Linux kernel, the most crucial component of many systems, we apply our method to a period of seven months before the release of Linux 5.4.
We find 29 commits that address 12 vulnerabilities.
For these vulnerabilities, our approach provides a temporal advantage of 2 to 179 days to design exploits before public disclosure takes place, and fixes are rolled out.

Established responsible disclosure approaches in open development processes are supposed to limit premature visibility of security vulnerabilities.
However, our approach shows that, instead, they open \emph{additional} possibilities to uncover such changes that thwart the very premise.
We conclude by discussing implications and partial countermeasures.

%% file: content.tex
\input{introduction}
\input{process}
\input{methodology}
\input{analysis}

\input{discussion}

\input{related}
\input{conclusion}
\input{ack}

%% file: introduction.tex
\section{Introduction}

% L1TF timeline:
% First public artefact on ML:
%   2018-08-14: http://lkml.iu.edu/hypermail/linux/kernel/1808.1/04003.html (for 4.18)
%   2018-08-14: Release Linus Mainline
%   2018-08-15: Release 4.18.1
%   2018-08-19: Debian 4.9.110-3+deb9u3

On 14 August 2018, a series of patches was integrated in Linux to provide mitigations for the Level 1 Terminal Fault (L1TF)~\cite{weisse2018foreshadowNG, vanbulck2018foreshadow} vulnerability\footnote{
See \href{https://git.kernel.org/pub/scm/linux/kernel/git/torvalds/linux.git/commit/?id=958f338e96f874a0d29442396d6adf9c1e17aa2d}{Linux commit 958f338e96} (hyperlink available in PDF).} -- a speculative execution attack with severe consequences that enable large scale data leakage across virtual machines on Intel-based cloud appliances.
While associated CVE entries were already filed in December 2017~\cite{l1tf-cve}, the vulnerability was embargoed until 14 August 2018~\cite{intel-l1tf-embargo} -- the same day of the disclosure and integration of the critical patches for Linux.
Unlike ordinary patches, these patches were---for obvious reasons---not discussed and developed on one of Linux's public communication channels (i.e., mailing lists) beforehand.

However, the fact that a patch was \emph{not} publicly discussed betrays it:
we will show that it is possible to detect such patches as soon as they enter a public repository.
This gives attackers valuable information advantage to design exploits.
For the aforementioned attack, it took another five days until the patches were integrated and rolled out by Debian 9,\footnote{See
\href{https://lists.debian.org/debian-security-announce/2018/msg00208.html}{the announcement} of
\href{https://salsa.debian.org/kernel-team/linux/commits/debian/4.9.110-3+deb9u3}{Debian kernel 4.9.110-3+deb9u3}.} a popular and wide-spread Linux distribution.

In this paper, we present a methodology to reverse engineer development processes:
We collect all publicly available development artefacts and map them against the software repository.
Using techniques to connect developer communication with repository entries~\cite{ramsauer2019list}, we are 
are able to uncover commits from non-public secret integration channels with high probability using semi-automatic methods. 
In this paper, and without loss of generality, we exercise the approach for mail-based development workflows (as used by the Linux kernel, QEMU, GCC, and many other projects), and show
how to systematically obtain \emph{off-list patches}: Code changes that were developed outside the official public lists.
Besides fixes for security vulnerabilities, we also find that there exist systematic channels to inject code into the Linux kernel while bypassing public discussion.
Our method provides two advantages for malicious attackers:
(a) it significantly reduces search efforts for fixes of security vulnerabilities, compared to fully manual investigation, and
(b) it provides temporal advantage for the design of attacks.
We claim following contributions:
\begin{compactitem}
    \item We present a method to systematically detect development process infringements in open-source projects that works as soon as commits arrive in repositories
    \item We detect and categorise different types of secret integration channels of the Linux kernel, such as \emph{bypass of development processes} or non-publicly discussed fixes for \emph{security vulnerabilities}
    \item We discuss methods that mitigate potential threats to open-source software ecosystems
\end{compactitem}

\paragraph{Outline}
The rest of this paper is structured as follows:
We first give an overview of the problem statement in Section~\ref{sec:secret}.
In Section~\ref{sec:process}, we provide a quick introduction to common open-source development practises as they are, for example, implemented by the Linux kernel community.
We then present our methodology of mapping development artefacts in Section~\ref{sec:methodology}.
In Sections~\ref{sec:analysis},  we run our analysis and evaluate a certain time window of the Linux kernel.
The discussion in Section~\ref{sec:discussion} focuses on potential threats to the ecosystem.
Section~\ref{sec:related} presents related work.
Finally, Section~\ref{sec:conclusion} concludes the paper and gives an outlook on future work.

\section{Secret Integration Channels}
\label{sec:secret}

The openness of the development processes is a key aspect of any open-source software (OSS) project:
Almost all development activities happen in public.
Since development artefacts (i.e., discussions or patch data on public mailing lists) are observable, this allows us to analyse the process in detail.

However, especially the development of fixes for critical security vulnerabilities intentionally happens behind closed scenes~\cite{linux-submitting-patches}.
After their disclosure, fixes silently appear as commits in the repository.
Nonetheless, those commits can not be assigned to any prior artefact that relates to its public pre-integration history.
Unless the vulnerability is explicitly announced or attracts medial attention, we disprove the common belief that patches typically drown in the noise of other commits in the repository.

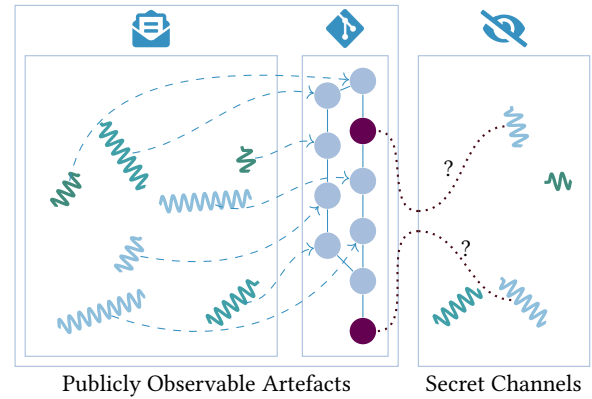
\begin{figure}
	\centering
	\input{img/mapping}
	\caption{Disclosing secret integration channels. On the left: artefacts on public channels (e.g., patches on mailing lists) are assigned to commits in the repository. On the right: Commits that lack assignable public artefacts arise from secret integration channels.}
	\label{fig:artefacts}
\end{figure}

Nevertheless, a full coverage of all public available development resources allows us to systematically exclude \emph{regular development noise} in order to separate it  from irregularities:
We mine for commits in repositories that come from secret integration channels---and detect them just-in-time to design exploits for vulnerabilities. % before fixes are rolled out.
Figure~\ref{fig:artefacts} illustrates the chase for missing links: we deduce that commits that can not be assigned to publicly observable artefacts must arise from secret integration channels.

Collaborative development tools, such as version control systems, bug trackers, continuous integration software or mailing lists~\cite{lanubile2010collaboration}, are central hubs in modern software projects.
Those tools gather collateral development artefacts and traces that directly relate to the the final result of the development process: an actual change of code in terms of a commit in a version control system, such as, for example, git~\cite{bird2009promises}.

In their everyday work, OSS developers present their patches to the public as part of the integration process.
On communication platforms, changes are discussed and reviewed before they are integrated by maintainers, trusted individuals that are authorised to commit changes to official resources~\cite{mauerer2013open}.
While several web-based collaboration systems aim to ease workflows of development processes~\cite{lanubile2010collaboration}, especially system software (e.g., operating systems, system level libraries or compilers) back mailing lists as their predominant tool of choice~\cite{bettenburg2015management, adams14-ll}.
Patches are wrapped in mails, sent to public lists and distributed to all subscribers.
Everyone is welcome to join the discussion in the mail thread and comment on the patch.
Later, the patch is picked up by a maintainer, who integrates it to their repository.
It is not untypical that maintainers fine-tune a patch before they apply it~\cite{bettenburg2015management, linux-submitting-patches}.

As one of the world's largest software undertakings~\cite{zhou2017scalability}, the Linux kernel is the core of a popular and wide-spread operating system and one of largest projects that follows this mail-based development model~\cite{linux-submitting-patches}.
More than 10,000 patches are integrated into each major release in eight week cadence.

Nevertheless, a rapidly changing code base, size and complexity inherently results in software defects that can lead to severe software vulnerabilities.
In 2019, 170 CVE entries were filed for all different versions and flavours of the Linux kernel, and many more potential security vulnerabilities have been fixed without CVE analysis and assignment~\cite{greg2019-cve}.
Unavoidably, the \emph{kernel community} has processes on managing critical vulnerabilities.

In contrast to regular development activities, vulnerabilities shall be reported to and discussed on private mailing lists~\cite{linux-submitting-patches}.
The rationale behind private \mbox{discussions} is the \emph{responsible disclosure} vulnerability disclosure model:
Software producers get the chance to provide fixes for vulnerabilities before they are publicly disclosed~\cite{cavusoglu2005emerging}.
Therefore, security mailing lists are closed-recipients lists to avoid early public attention.
Only carefully selected and trusted individuals have permission to join those lists.
Security lists are used for coordination, and to setup private communication between reporters and affected subsystems.
They can also be used to develop the actual fixes for the issues~\cite{greg2020-private}.

Eventually, when the fix is in its final state, it is released for all affected version of the kernel that are supported by the community:
This leaves the first publicly visible footprint of the vulnerability: the patch(es) in the repository.
Yet, it misses a link to a publicly observable artefact.

%% file: img/mapping.tex
\begin{tikzpicture}[scale=0.67]
% \definecolor{col1}{RGB}{241,238,246}
% \definecolor{col2}{RGB}{212,185,218}
% \definecolor{col3}{RGB}{201,148,199}
% \definecolor{col4}{RGB}{223,101,176}
% \definecolor{col5}{RGB}{231,041,138}
% \definecolor{col6}{RGB}{206,018,086}
% \definecolor{col7}{RGB}{145,000,063}

\definecolor{col1}{RGB}{246,239,247}
\definecolor{col2}{RGB}{208,209,230}
\definecolor{col3}{RGB}{166,189,219}
\definecolor{col4}{RGB}{103,169,207}
\definecolor{col5}{RGB}{54,144,192}
\definecolor{col6}{RGB}{2,129,138}
\definecolor{col7}{RGB}{1,100,80}

\definecolor{red1}{RGB}{100,1,80}
\definecolor{red2}{RGB}{70,10,20}

% \definecolor{col1}{RGB}{240,249,232}
% \definecolor{col2}{RGB}{204,235,197}
% \definecolor{col3}{RGB}{168,221,181}
% \definecolor{col4}{RGB}{123,204,196}
% \definecolor{col5}{RGB}{78,179,211}
% \definecolor{col6}{RGB}{43,140,190}
% \definecolor{col7}{RGB}{8,88,158}

	\newcommand\commit[1][]{\node[commit, #1]}
	\newcommand\githist[2]{\draw[draw=col5] (#1) -- (#2)}
	
	\newcommand\thread[3]{
	    \draw [thread,very thick, draw=#3!75] (#1) -- +(#2) node (m#1) [midway] {};
	}
	\newcommand{\threadlink}[4]{\draw [draw=col5,dashed,->] (#1) to [out=#2, in=#3] (#4)}
	
    \newcommand{\backrectangle}[2]{\draw[draw=col3] (#1) rectangle (#2)}
    \newcommand{\rectanglecoordinates}[5]{
        \coordinate (#1bl) at (#2 |- #3);
        \coordinate (#1br) at (#4 |- #3);

        \coordinate (#1tl) at (#2 |- #5);
        \coordinate (#1tr) at (#4 |- #5);
        
        \coordinate (#1ml) at ($(#1bl)!.5!(#1tl)$);
        
        \coordinate (#1bm) at ($(#1bl)!.5!(#1br)$);
        \coordinate (#1tm) at ($(#1tl)!.5!(#1tr)$);
        
        \begin{scope}[on background layer]
        \backrectangle{#1bl}{#1tr};
        \end{scope}
    }

%%%%%%%%%%%%%%%%%%%%%%%%%%%%%%%%%%%%%%%%%%%%%%%%%%%%%

	\tikzset{
		thread/.style={decorate, decoration={snake, amplitude=3pt, segment length=4pt}, draw=red},
		commit/.style={circle, fill=col3, minimum size=10pt, inner sep=0pt, node distance=.3},
		threadnode/.style={node distance=0.5em}
	}

% 	\commit[fill=col5] (0) at (0, 0) {0};
% 	\commit[above = of 0] (1) {1};
% 	\commit[above = of 1] (2) {2};
% 	\commit[above = of 2] (3) {3};
% 	\commit[above = of 3, fill=col5] (4) {4};
% 	\commit[above = of 4] (5) {5};
% 	\commit[above left=of 1] (7) {7};
% 	\commit[above = of 7] (8) {8};
% 	\commit[above = of 8] (9) {9};
% 	\commit[above = of 9] (10) {10};
	
	\commit[fill=red1] (0) at (0, 0) {};
	\commit[above = of 0] (1) {};
	\commit[above = of 1] (2) {};
	\commit[above = of 2] (3) {};
	\commit[above = of 3, fill=red1] (4) {};
	\commit[above = of 4] (5) {};
	\commit[above left=of 1] (7) {};
	\commit[above = of 7] (8) {};
	\commit[above = of 8] (9) {};
	\commit[above = of 9] (10) {};

	\foreach \from/\to in {0/1, 1/2, 2/3, 3/4, 4/5,
			       7/8, 8/9, 9/10, 5/10, 7/1}
		\githist{\from}{\to};

	\node (common_thread_node) at ($(0) - (6.2,0)$) {};
	\node (secret_thread_node) at ($(0) + (1.3,0)$) {};

	\foreach \name/\offx/\offy/\length/\color in
	{t1/ 0.0/ 0.0/ 20:  2.0/col4,
	 t2/ 3.0/ 0.0/ 41:  1.5/col6,
	 t4/ 1.3/ 1.0/ 60:  1.0/col4,
	 t6/ 0.0/ 2.3/ 55:  1.0/col7,
	 t7/ 2.0/ 2.5/  4:  2.0/col4,
	 t8/ 2.0/ 2.7/125: 1.75/col6,
	 t9/ 4.0/ 3.0/107:  0.7/col7}
	{
		\node (\name) at ($(common_thread_node) + (\offx,\offy)$) {};
		\thread{\name}{\length}{\color};
	}
	
	\foreach \name/\offx/\offy/\length/\color in
	{s1/ 1.9/ 3.5/105:  1.0/col4,
	 s2/ 0.0/ 0.0/ 40:  1.3/col6,
	 s3/ 3.0/3.0/  184: 0.7/col7,
	 s4/ 2.5/ 0.0/135:  1.5/col4}
	{
		\node (\name) at ($(secret_thread_node) + (\offx,\offy)$) {};
		\thread{\name}{\length}{\color};
	}

	\coordinate (lbl) at ($(t1) - (.7, .7)$);
	\coordinate (ltr) at ($(5)+(.7, 1.5)$);
	
	\coordinate (repbl) at ($(7|-0) - (.5,.5)$);
	\coordinate (reptr) at ($(5) + (.5,.5)$);
	
	\coordinate (mailbl) at ($(t1) - (.5,.5)$);
	\coordinate (mailtr) at ($(t9 |- 5) + (.5,.5)$);
    
	\coordinate (rbl) at ($(s2 |- 0) - (.2, .7)$);
	\coordinate (rtr) at ($(s3 |- 5)+(.2, .5)$);

	\rectanglecoordinates{left}{lbl}{lbl}{ltr}{ltr}
	\rectanglecoordinates{repo}{repbl}{repbl}{reptr}{reptr}
	\rectanglecoordinates{mailing}{mailbl}{mailbl}{mailtr}{mailtr}
	\rectanglecoordinates{right}{rbl}{rbl}{rtr}{rtr}

	\node[below] at (leftbm) {Publicly Observable Artefacts};

% 	\node[above] at (mailingtm) {Patches on Mailing Lists};
% 	\node[above] at (repotm) {Repository};

	\node[below] at  ($(rightbm) - (0.0, 0)$){Secret Channels};
% 	\node[above] at  (righttm){Secret channels\phantom{y}};
	
	\node[above] at (mailingtm) {\includegraphics[height=.5cm]{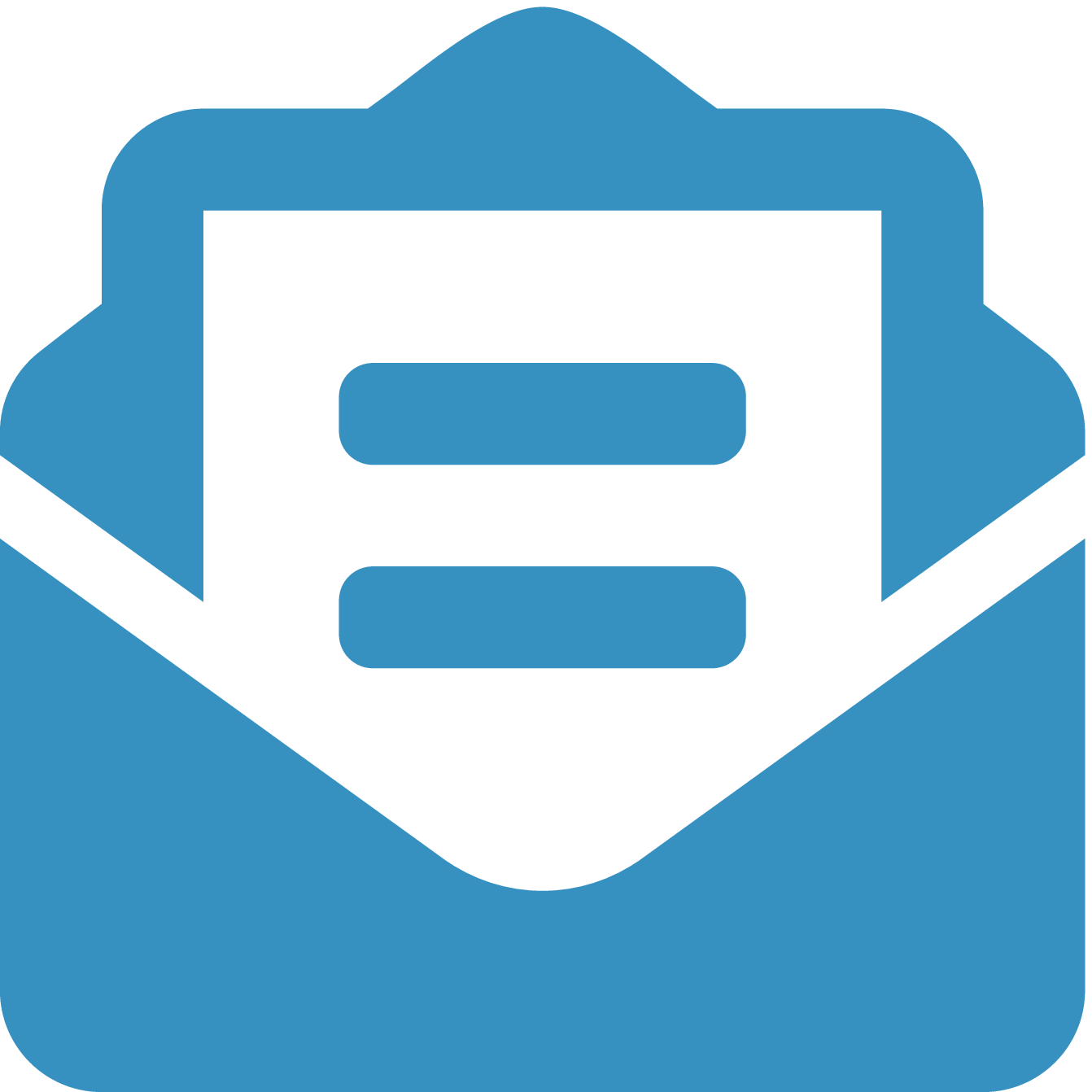}};
	\node[above] at (repotm) {\includegraphics[height=.5cm]{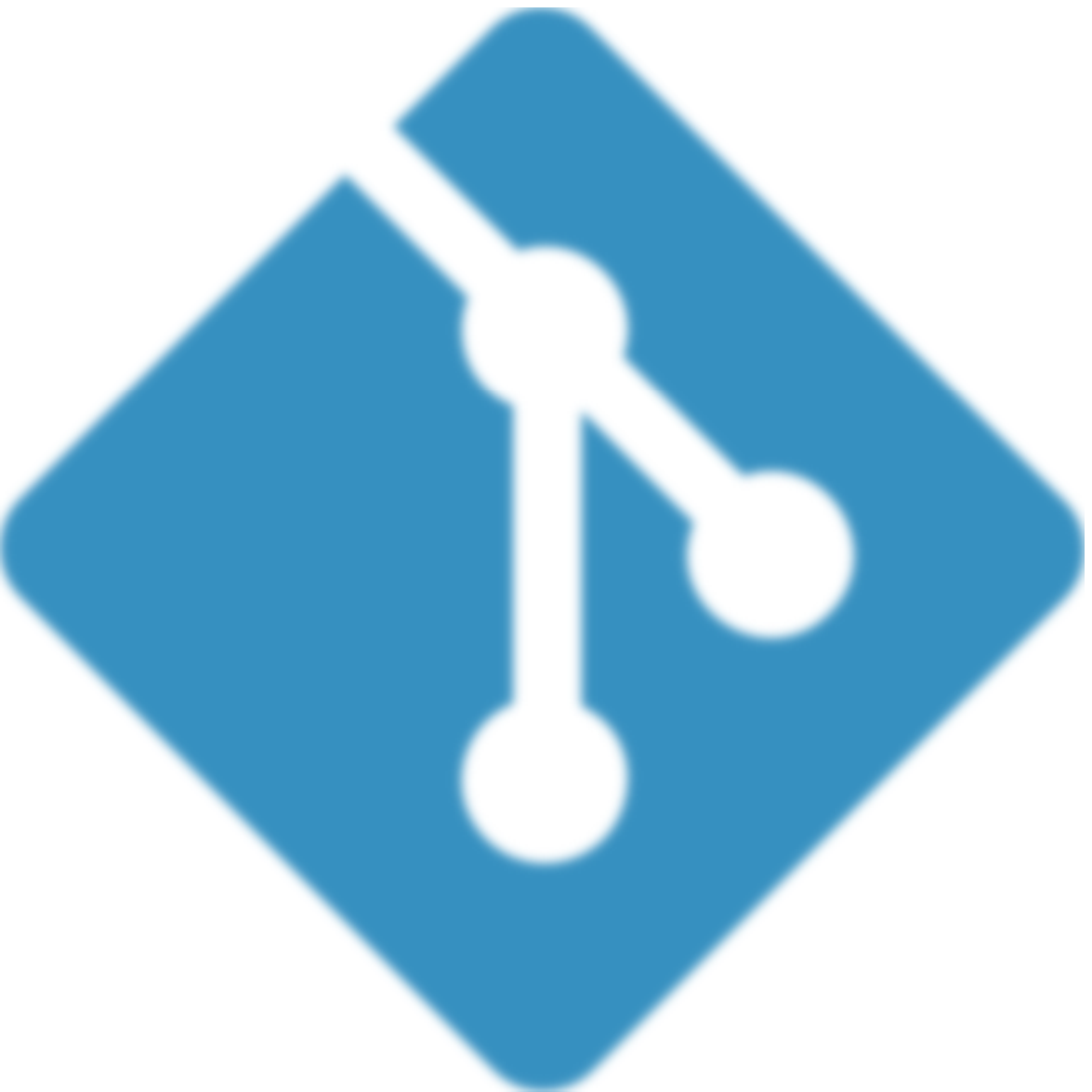}};
	\node[above] at (righttm) {\includegraphics[height=.5cm]{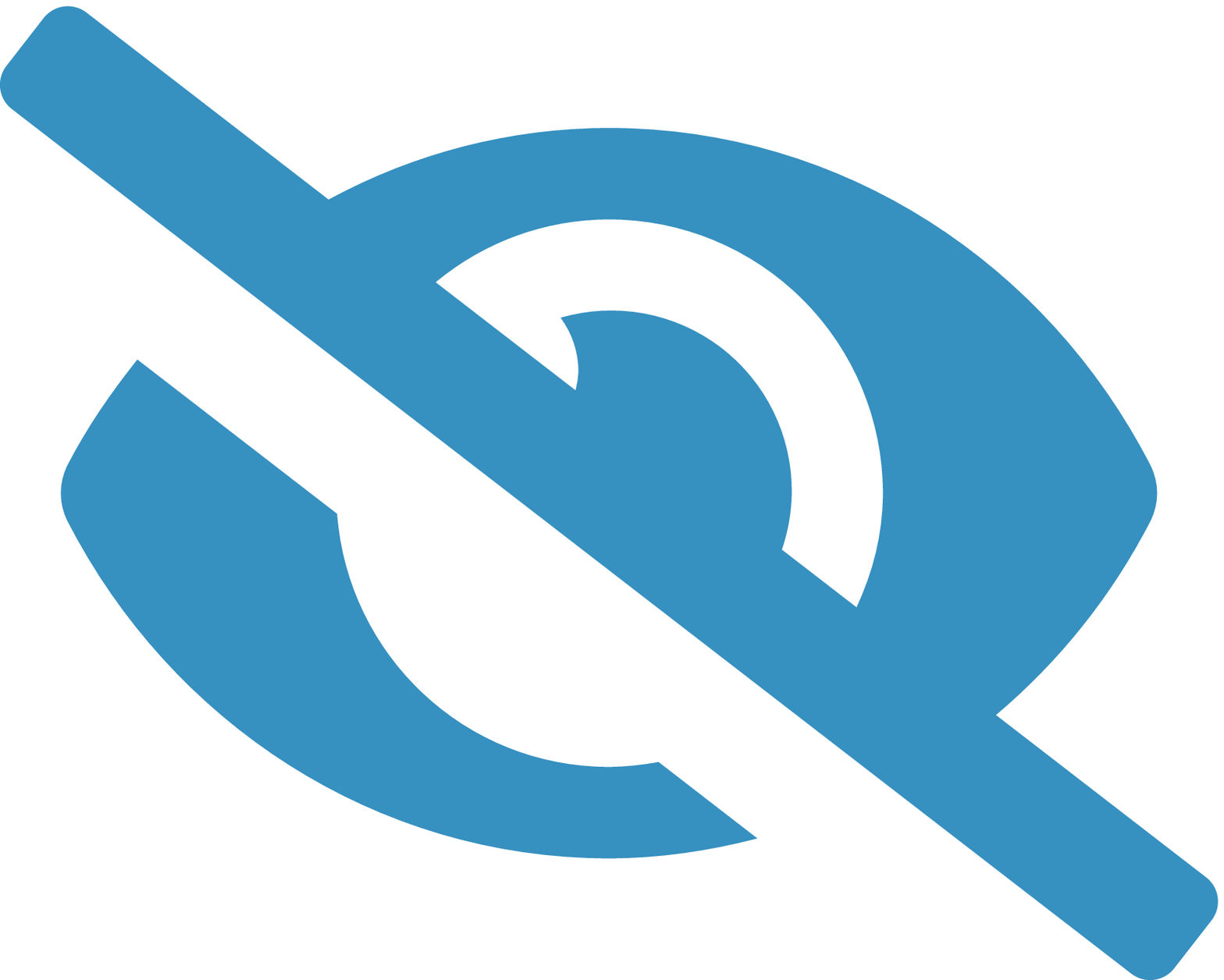}};
	
% 	\draw[draw=col3,decorate,decoration={brace,amplitude=10pt,mirror}] ([xshift=-2pt]righttl) -- ([xshift=-2pt]rbl) 
% 	node (brace) [black,midway,xshift=-10pt,inner sep=0pt] {};
	
	\threadlink{mt1}{340}{240}{2};
	\threadlink{mt2}{  0}{190}{7};
	\threadlink{mt4}{350}{230}{8};
	\threadlink{mt6}{ 79}{175}{5};
	\threadlink{mt7}{340}{170}{3};
	\threadlink{mt8}{ 10}{170}{10};
	\threadlink{mt9}{  5}{166}{9};

	\draw [draw=red2,thick, dotted] (rightml) to [out=180, in=0] (4);
	\draw [draw=red2,thick, dotted] (rightml) to [out=0, in=180] node [left, midway] {?} (ms1);

	\draw [draw=red2,thick, dotted] ($(rightml) + (0, -0.4)$) to [out=180, in=0] (0);
	\draw [draw=red2,thick, dotted] ($(rightml) + (0, -0.4)$) to [out=0, in=180] node [above, midway] {?} (ms4);
\end{tikzpicture}

%% file: process.tex
\section{Linux Kernel Development Process}
\label{sec:process}
\epigraph{Given enough eyeballs, all bugs are shallow}{\textit{Linus' Law -- by Eric S. Raymond}~\cite{raymond1999cathedral}}

\noindent  This section gives a brief overview of the Linux kernel development process (LKDP).
One peculiarity of the LKDP is the large number of contributors (thousands per year) and
participants, which lead to the well-known hypothesised connection given above between the
decreasing difficulty of detecting bugs with an increasing number of reviewers. Since we abuse the principle
to detect patches that have seemingly \emph{not} receive sufficient public attention, it is pertinent to recapture key characteristics of the development process that are  relevant for our approach.

\subsection{Core Characteristics}
Development of the Linux kernel proceeds in two-phase cycles: New code and features are merged during a two-week long \emph{merge window}, which is followed by a two-month long \emph{stabilisation window}~\cite{linux-process}.
This leads to development cycles of approximately 2.5 months between two major releases. More than 10,000 patches are integrated  in each cycle into Linus Torvalds' (the project owner's) git tree, which is commonly called \textit{Linux mainline}.
Before code changes (\emph{patches}) are integrated into mainline, they must have been discussed on a public mailing list.
This is demanded by the submission guidelines of Linux, and is intended to ascertain good code quality~\cite{meneely2009secure}.
Because of scalability, availability, robustness, and simplicity, many low-level system software components prefer mail-based communication over using web-based technologies~\cite{greg2016-mail}.
Other communication channels channels need not be considered for our purposes.

Similar to a commit in a repository, an email encapsulates a patch that contains a commit message, an informal description of the changes, and a \emph{diff} that specifies insertions and deletions
of code---relative to a specific code base.
Typically, larger logical changes are split into multiple small patches.
This gives a \emph{patch series} whose elements are tied together by a \emph{cover letter}.
Cover letters give an informal, higher-level overview of the series.
Together with the proper patches, it is sent as a mail thread to maintainers and the corresponding list(s) of the affected subsystem(s) of the project.

Everyone can join the discussion of patches as lists are usually unmoderated.
Maintainers who are responsible for the list to which the series is posted,
or for a subsystem that the patch addresses, eventually
  (a) refuse the patch,
  (b) ask for further refinement of the patch,
  (c) pick up the patch and commit it to their maintainer tree.
Maintainer trees are staging points before code changes are finally integrated mainline.
It is not unusual that (b) is repeated over several iterations until the patch series is deemed acceptable for merging.

Because of the massive number of emails and patches, the Linux  kernel currently utilises over 200 different mailing lists that are logically partitioned by topic or subsystem.
On average, an email is received by one of those lists every 20 seconds.

Maintainers are organised in a semi-formal hierarchy~\cite{Mauerer:2010}.
During a \emph{merge window}, maintainers ask hierarchically higher-level maintainers to \emph{pull their changes}, which is possible in two ways:
Either by picking up and integrating patch data from mailing lists, or by \emph{pulling} code from repositories. Once the top-level maintainer Linus Torvalds pulls and publishes changes, they become part of Linux mainline.

\subsection{Lifecycle Management}

\begin{figure*}
	\input{img/backport}
	\caption{Linux development timeline: Mainline, stable trees and distribution trees are supported in parallel. Typically, fixes for vulnerabilities are first fixed in
	mainline (cf.~Vulnerability 1) and on a stable tree, before they are ported back by distributions. In rare cases (cf.~Vulnerability 2), patches appear in distributions before
	they are published in mainline.}
	\label{fig:backport}
\end{figure*}
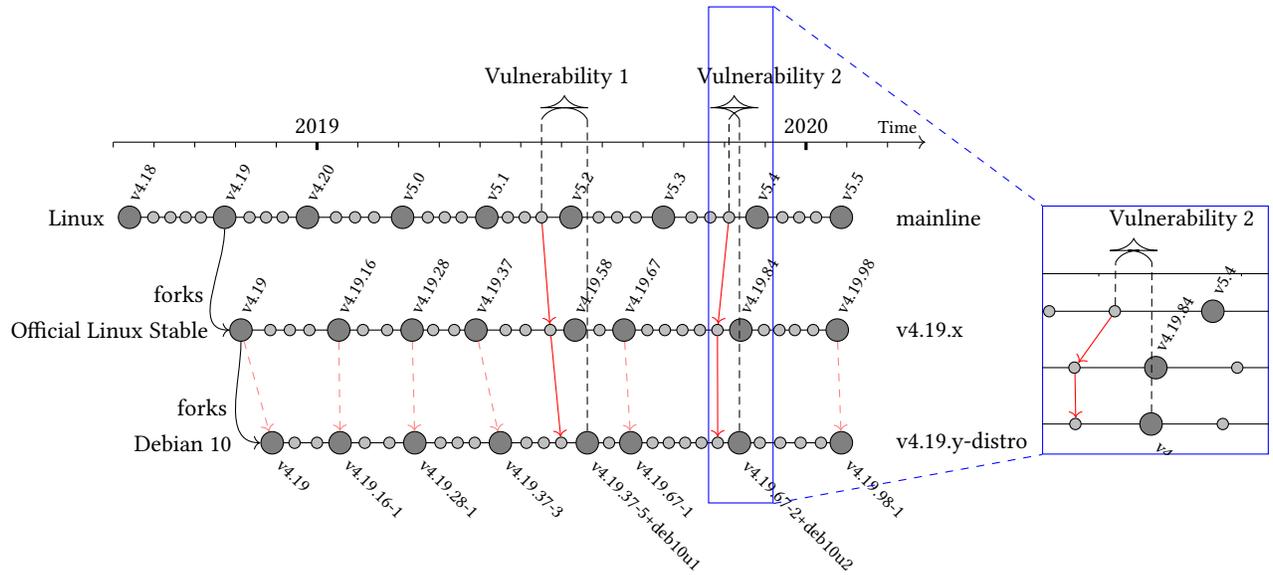

The latest release of Linux is called the \emph{stable tree}, and is actively supported with bug-fixes until the next mainline release is cut, and becomes the new stable tree.
Additionally, the Linux kernel community supports several versions of the kernel in parallel~\cite{kroahhartman2008linux} that are referred to as long-term support (LTS) versions.
They are based on selected stable trees, and  receive official support for up to six years.
Figure~\ref{fig:backport} illustrates the parallel development of mainline Linux and the maintenance of LTS versions.

Linux distributions and vendors usually choose LTS versions as the basis of their kernel (which may additionally contain a substantial amount of added drivers, domain-specific features, and many other additional elements), since they provide a stable and reliable base that will not be subjected to invasive changes (e.g., API changes) during their lifetime.
New features are only accepted mainline.
Stable and LTS trees may only receive stabilisation patches, bug fixes, or fixes for vulnerabilities.

In case patches to LTS versions are also relevant for mainline, they must be, by the \emph{upstream first} convention, integrated in mainline before they are ported back to stable releases.
After their release, distributions pick patches from stable versions and apply them to their own kernel repository.
From a temporal perspective, the typical pathway of a bug fix is mainline$\rightarrow$stable$\rightarrow$distribution.

\subsection{Exceptional Vulnerability Handling}

The aforementioned public review and integration process allows for an exception when 
fixes for security vulnerabilities must be handled.
The Linux kernel is a key software component of a large class of machines from embedded
industrial control appliances to cloud computing servers.
Consequently, the Linux kernel community has established
standard procedures for responsible disclosure~\cite{christey2002responsible, frei2010modeling}.

Linux submission guidelines encourage developers to report exploitable security bugs to the non-public security team mailing list \href{security@kernel.org}{security@kernel.org}:
``For severe bugs, a short embargo may be considered to allow distributors to get the patch out to users; in such cases, obviously, the patch should not be sent to any public lists.''~\cite{linux-submitting-patches}

Similar to the regular public development process, patches for vulnerabilities are iteratively discussed, reviewed and refined -- but all related conversations take either place in private email conversations or on closed lists.
Once participants agree on a fix~\cite{greg2020-private}, or after embargoes are expired, the majority of fixes follow the same procedures as bugs:
Patches for mainline and affected stable versions are published at the same time, before they are integrated into distribution repositories.
Figure~\ref{fig:backport} (Vulnerability 1) illustrates the temporal process of a typical vulnerability.
There is a second type of coordinated disclosure for severe vulnerabilities that we discuss in Section~\ref{sec:discussion}.

%% file: img/backport.tex
\ifdef{\isstandalone}{
	\setlength{\linewidth}{469.75499pt}
}

\begin{tikzpicture}
        \newdimen\curx
        \newlength{\xaxiswidthpt}

        \newcommand{\rectanglecoordinates}[7]{
            \coordinate (#1bl) at (#2 |- #3);
            \coordinate (#1tl) at ($(#1bl)+(0,#4)$);
            \coordinate (#1ml) at ($(#1bl)!.5!(#1tl)$);

            \coordinate (#1br) at (#5 |- #1bl);
            \coordinate (#1mr) at (#5 |- #1ml);
            \coordinate (#1tr) at (#5 |- #1tl);

            \coordinate (#1bm) at ($(#1bl)!.5!(#1br)$);
            \coordinate (#1tm) at ($(#1tl)!.5!(#1tr)$);
            \coordinate (#1mm) at ($(#1ml)!.5!(#1mr)$);

            %\draw (#1bl) rectangle (#1tr);

            \node (#1westtext) [anchor=east] at (#1ml) {#6};
            \node (#1easttext)[anchor=west] at ($(#1mr) - (0.5, 0)$) {#7};
        }

        \newcommand{\mycircle}[3]{
            \node[circle,draw=black,fill=red,minimum size=5pt,inner sep=0pt] (#1) at (#2 |- #3) {}
        }
        
        \newcommand{\spyon}[7]{
        %inputs: 
        % #1{spy-on coordinate} eg. {(0,3)}, testcoordinate
        % #2{spy-in coordinate}
        % #3{spy-in width}
        % #4{spy-in height}
        % #5{spy-xscale} 
        % #6{spy-yscale}
        % #7{image}
            
            \pgfmathsetmacro{\myrand}{random(1,1000000)};
            \def\spydrawcolor{blue}
            \node[draw=\spydrawcolor,inner sep = 0,
            minimum width={#3/#5},minimum height={#4/#6}] (spyONnode\myrand) at #1 {};
            %% spy in node
            \node[,draw=\spydrawcolor, inner sep = 0,
            minimum height=#4,minimum width=#3] (spyINnode\myrand) at #2 {};
        
            \begin{scope}
                \clip (spyINnode\myrand.south west) rectangle (spyINnode\myrand.north east);
                \begin{scope}[
                    shift={($(spyINnode\myrand)-#5*(spyONnode\myrand |- {(0,0)})-#6*(spyONnode\myrand -| {(0,0)})$)},
                    xscale=#5,
                    yscale=#6
                ]
                    #7
                \end{scope}
            \end{scope}
            % \draw[draw=\spydrawcolor,dashed] (spyONnode\myrand) -- (spyINnode\myrand);
            
            \draw[draw=\spydrawcolor,dashed] (spyONnode\myrand.north east) -- (spyINnode\myrand.north west);
            
            \draw[draw=\spydrawcolor,dashed] (spyONnode\myrand.south east) -- (spyINnode\myrand.south west);
        }
        
        \newcommand\release[3]{
            \IfEndWith{#1}{4.19.0}{
            \node [fill=gray, draw, circle, minimum size=0.25cm, label={[rotate=#2]#3:\footnotesize v4.19}]
            }{
            \node [fill=gray, draw, circle, minimum size=0.25cm, label={[rotate=#2]#3:\footnotesize #1}]
            }
        }
        
        \newcommand\commit[0]{
            \node [draw, circle, inner sep=0pt, minimum size=0.15cm, fill=lightgray]
        }
        
        \newcommand\embrace[3]{
            \coordinate (uvuln1#3) at ($(#1) + (0, 4.3)$);
            \coordinate (uvuln2#3) at ($(#2) + (0, 1.3)$);
            
            \draw[densely dashed] (#1) -- (uvuln1#3);
            \draw[densely dashed] (#2) -- (uvuln2#3);
        }
        
        \newcommand\port{
        \draw[->, draw=red!50, dashed]
        }
        
        \newcommand\vuln{
            \draw[->, draw=red]
        }

        \newcommand{\axeh}{6.25}
        \newcommand{\bheight}{.5}

        \setlength{\xaxiswidthpt}{\linewidth - \widthof{v4.19.y-distroLinux} -4.5cm}

        \newcounter{monthStart}\setcounter{monthStart}{7}
        \newcounter{imonth}\setcounter{imonth}{0}
        \newcommand{\yearStart}{2018}
        \newcommand{\yearEnd}{2020}
        \pgfmathsetmacro\years{\yearEnd - \yearStart +1}
        \pgfmathsetmacro\linewidthcm{\xaxiswidthpt /28.35 }
        \pgfmathsetmacro\monthSpacing{\linewidthcm / 20} % 20 months are shown in sum
        \pgfmathsetmacro\daySpacing{\monthSpacing / 30}

        \def\pic{
        %%%%%%%%%%%%%%%%%%%%%%%%%%%%%%%%%%%%%%%%%%%%%%%%%%%%%%%%%
        %%%%%%%%%%%%%%%%%%%%%%%%%%%%%%%%%%%%%%%%%%%%%%%%%%%%%%%%%
        %% NO COMMANDS IN HERE %%%%%%%%%%%%%%%%%%%%%%%%%%%%%%%%%%
                    \coordinate (start) at (0, \axeh);
        \coordinate (end) at (\xaxiswidthpt, \axeh);

        \coordinate (h2) at (0,2);
        \coordinate (h3) at (0,3.5);
        \coordinate (h4) at (0,5);

        \foreach \x in {\yearStart,...,\yearEnd}{
                \coordinate (\x) at ($(start) + \theimonth*(\monthSpacing, 0)$);
                \ifthenelse{\value{monthStart}=1}{
                	\node[above]  at (\x) {\x};
                	\draw[very thick] (\x) -- ($(\x) + (0,-3pt)$);
			\refstepcounter{imonth}
            	}{}

                \foreach \xi [count=\i from 1] in {\themonthStart,...,11}{
                        \coordinate (\x_\xi) at ($(start) + \theimonth*(\monthSpacing,0)$);
                        \pgfextractx{\curx}{\pgfpointanchor{\x_\xi}{center}}

                        \ifthenelse{\curx < \xaxiswidthpt}{
                        	\draw (\x_\xi) -- ($(\x_\xi) + (0,-2pt)$);
                        }{}
                        \refstepcounter{imonth}
                }
                \setcounter{monthStart}{1}%
        }
        \setcounter{imonth}{0}
        \setcounter{monthStart}{7}
        \rectanglecoordinates{m}{start}{h4}{\bheight}{end}{Linux}{mainline}

        \coordinate (_mv418) at ($(2018_7|-mmm) + \daySpacing*(12, 0)$);
        \coordinate (_mv55) at ($(2020|-mmm) + \daySpacing*(26, 0)$);
        \draw [-] (_mv418) -- (_mv55);

        \foreach \major/\minor/\month/\day in {
            %v4/15/2018/27,
            %v4/16/2018_3/1,
            %v4/17/2018_5/3,
            v4/18/2018_7/12,
            v4/19/2018_9/22,
            v4/20/2018_11/23,
            v5/0/2019_2/3,
            v5/1/2019_4/5,
            v5/2/2019_6/7,
            v5/3/2019_8/15,
            v5/4/2019_10/24,
            v5/5/2020/26}{
            \release{\major.\minor}{60}{15} (m\major\minor) at ($(\month|-mmm) + \daySpacing*(\day, 0)$) {};
        }
        
        \foreach \name/\start/\stop/\scale in {
            %/mv415/mv416/0.3,
            %/mv415/mv416/0.5,
            %/mv415/mv416/0.7,
            %/mv416/mv417/0.3,
            %/mv416/mv417/0.5,
            %/mv416/mv417/0.7,
            %/mv417/mv418/0.3,
            %/mv417/mv418/0.5,
            %/mv417/mv418/0.7,
            /mv418/mv419/0.25,
            /mv418/mv419/0.43,
            /mv418/mv419/0.59,
            /mv418/mv419/0.75,
            /mv419/mv420/0.3,
            /mv419/mv420/0.5,
            /mv419/mv420/0.7,
            /mv420/mv50/0.3,
            /mv420/mv50/0.5,
            /mv420/mv50/0.7,
            /mv50/mv51/0.3,
            /mv50/mv51/0.5,
            /mv50/mv51/0.7,
            /mv51/mv52/0.25,
            /mv51/mv52/0.45,
            mvuln1/mv51/mv52/0.65,
            /mv52/mv53/0.3,
            /mv52/mv53/0.5,
            /mv52/mv53/0.7,
            /mv53/mv54/0.3,
            /mv53/mv54/0.5,
            mvuln2/mv53/mv54/0.7,
            /mv54/mv55/0.3,
            /mv54/mv55/0.5,
            /mv54/mv55/0.7}{
            \commit (\name) at ($(\start)!\scale!(\stop)$) {};
        }
        
        \rectanglecoordinates{s}{{mv419.east}}{h3}{\bheight}{end}{Official Linux Stable\phantom{M}}{v4.19.x}
        
        \coordinate (_sv4190) at ($(2018_10|-smm) + \daySpacing*(4, 0)$);
        \coordinate (_sv41998) at ($(2020|-smm) + \daySpacing*(23, 0)$);
        \draw [-] (_sv4190) -- (_sv41998);
        
        \foreach \patchlevel/\month/\day in {
            0/2018_10/4,
            16/2019/16,
            28/2019_2/10,
            37/2019_3/27,
            58/2019_6/10,
            67/2019_7/16,
            %80/2019_9/17,
            84/2019_10/12,
            %87/2019_11/1,
            98/2020/23}{
            \release{v4.19.\patchlevel}{60}{15} (sv419\patchlevel) at ($(\month|-smm) + \daySpacing*(\day, 0)$) {};
        }
        
        \foreach \name/\start/\stop/\scale in {
            /sv4190/sv41916/0.3,
            /sv4190/sv41916/0.5,
            /sv4190/sv41916/0.7,
            /sv41916/sv41928/0.33,
            /sv41916/sv41928/0.66,
            /sv41928/sv41937/0.33,
            /sv41928/sv41937/0.66,
            /sv41937/sv41958/0.3,
            /sv41937/sv41958/0.5,
            svuln1/sv41937/sv41958/0.75,
            /sv41958/sv41967/0.5,
            /sv41967/sv41984/0.2,
            /sv41967/sv41984/0.35,
            /sv41967/sv41984/0.5,
            /sv41967/sv41984/0.65,
            svuln2/sv41967/sv41984/0.8,
            /sv41984/sv41998/0.24,
            /sv41984/sv41998/0.40,
            /sv41984/sv41998/0.56,
            /sv41984/sv41998/0.75}{
            \commit (\name) at ($(\start)!\scale!(\stop)$) {};
        }
        
        \coordinate (foo) at ($(sbl) + (1em, 0)$);
        \rectanglecoordinates{d}{foo}{h2}{\bheight}{end}{Debian 10\phantom{M}}{v4.19.y-distro}
        
        \coordinate (_dv4190) at ($(2018_10|-dmm) + \daySpacing*(27, 0)$);
        \coordinate (_dv41998-1) at ($(2020|-dmm) + \daySpacing*(26, 0)$);
        \draw [-] (_dv4190) -- (_dv41998-1);
        
        \foreach \patchlevel/\month/\day in {
            0/2018_10/27,
            %13-1/2018_11/30,
            16-1/2019/17,
            28-1/2019_2/12,
            37-3/2019_4/15,
            37-5+deb10u1/2019_6/19,
            67-1/2019_7/21,
            67-2+deb10u2/2019_10/11,
            %87-1/2019_11/03,
            98-1/2020/26}{
            \release{v4.19.\patchlevel}{-45}{-15} (dv419\patchlevel) at ($(\month|-dmm) + \daySpacing*(\day, 0)$)
            {};
        }

        \foreach \name/\start/\stop/\scale in {
            /dv4190/dv41916-1/0.33,
            /dv4190/dv41916-1/0.66,
            /dv41916-1/dv41928-1/0.33,
            /dv41916-1/dv41928-1/0.66,
            /dv41928-1/dv41937-3/0.3,
            /dv41928-1/dv41937-3/0.5,
            /dv41928-1/dv41937-3/0.7,
            /dv41937-3/dv41937-5+deb10u1/0.3,
            /dv41937-3/dv41937-5+deb10u1/0.5,
            dvuln1/dv41937-3/dv41937-5+deb10u1/0.7,
            /dv41937-5+deb10u1/dv41967-1/0.5,
            /dv41967-1/dv41967-2+deb10u2/0.2,
            /dv41967-1/dv41967-2+deb10u2/0.35,
            /dv41967-1/dv41967-2+deb10u2/0.5,
            /dv41967-1/dv41967-2+deb10u2/0.65,
            dvuln2/dv41967-1/dv41967-2+deb10u2/0.8,
            /dv41967-2+deb10u2/dv41998-1/0.2,
            /dv41967-2+deb10u2/dv41998-1/0.4,
            /dv41967-2+deb10u2/dv41998-1/0.6,
            /dv41967-2+deb10u2/dv41998-1/0.8}{
            \commit (\name) at ($(\start)!\scale!(\stop)$) {};
        }

        \draw[->] (start) to (end);
        \node[anchor=south east] (time) at (end) {\footnotesize Time};
        
        \draw[->] (mv419) to [out=-90, in=180] node[anchor=east, midway] {forks} (sv4190);
        \draw[->] (sv4190) to [out=-90, in=180] node[anchor=east, midway] {forks} (dv4190);
        
        \port (sv4190) -- (dv4190);
        \port (sv41916) -- (dv41916-1);
        \port (sv41928) -- (dv41928-1);
        \port (sv41937) -- (dv41937-3);
        \port (sv41967) -- (dv41967-1);
        \port (sv41998) -- (dv41998-1);
        
        \vuln (mvuln1) -- (svuln1);
        \vuln (svuln1) -- (dvuln1);
        
        \vuln (mvuln2) -- (svuln2);
        \vuln (svuln2) -- (dvuln2);

        \embrace{dv41967-2+deb10u2}{mvuln2}{1}
        \embrace{dv41937-5+deb10u1}{mvuln1}{2}
        
        \draw[decorate,decoration={brace,amplitude=9pt}] (uvuln21) -- (uvuln11) node[above,yshift=0.3cm,xshift=0.4cm] {Vulnerability 2};
        \draw[decorate,decoration={brace,amplitude=9pt}] (uvuln22) -- (uvuln12) node[above,xshift=-0.4cm,yshift=0.3cm] {Vulnerability 1};

        %% NO COMMANDS UNTIL HERE %%%%%%%%%%%%%%%%%%%%%%%%%%%%%%%
        %%%%%%%%%%%%%%%%%%%%%%%%%%%%%%%%%%%%%%%%%%%%%%%%%%%%%%%%%
        %%%%%%%%%%%%%%%%%%%%%%%%%%%%%%%%%%%%%%%%%%%%%%%%%%%%%%%%%
        }
        
        %This actually draws all of the above stuff
        \pic

        \spyon{($(sv41984)+(0,1)$)}{($(seasttext)+(3,0)$)}{3cm}{3.3cm}{3.5}{.5}{\pic}
\end{tikzpicture}

%% file: methodology.tex
\section{Methodology}
\label{sec:methodology}
The basic idea of our approach is to mine public Linux repositories for patches that have not been discussed on any public mailing list.
Previous work~\cite{adams13-makeit, adams14-ll} associates ongoing development in email threads to commits in repositories such that pre-integration histories of changes can be uncovered.
For our purposes, we basically need to reverse the question, and find commits in repositories that do \emph{not} enjoy any traceable pre-integration history.

Temporally, a patch should first appear on a public mailing list before it can be found in the repository.
Hence, any new commits in the repository that can not be assigned to emails were integrated through non-public integration channels.

The nature of OSS allows for collecting all required development artefacts.
However, assigning patches on mailing lists to commits in repositories is a non-trivial task caused by the lack of reliable, machine-encoded provenance information.

\subsection{Information Processing}
Commits in repositories are identified by their unique commit hash, and
patches on mailing lists are identified by their unique \emph{Message-ID}.
Yet, the mapping of Message-IDs to commit hashes is lost during the manual integration process of maintainers~\cite{adams14-ll, bird2009promises}.

The kernel community is aware of the gap of code traceability, which is a frequent subject of meta-discussions~\cite{khan2019-ksummit-discussion, anderson2019-ksummit-discussion} on the improvement of the development process.
The issue is unsolved by the community at the time of writing.

Recent publications from the software-engineering community attempt to use mining techniques to reconstruct the pre-integration history of software projects in ex-post analyses (see, for instance, Refs.~\cite{ramsauer2019list, adams14-ll}).
Simple textual comparison fails to recover the history with high accuracy~\cite{adams14-ll}, as patches on mailing lists may significantly differ from their counterparts in the repository:
Maintainers rewrite commit messages, add additional changes to the code, move code, or apply patches against a different state if the code base~\cite{bird2009promises, adams14-ll}, which can (slightly) change the content of the committed itself.
The initial version of a patch can significantly vary from further revisions or from its final state in the repository.

While several approaches to rate similarity rating of patches have been devised~\cite{adams14-ll, adams13-makeit, ramsauer2019list}, each
provides a similarity score sim for a pair of patches.
For the comparison of patches, we do, in the first place, not differentiate between patches as commits in repositories or patches as mails on mailing lists.
Let $\mathcal{M}$ be the set of patches on mailing lists and
$\mathcal{C}$ be the set of commits in the repository. Let further $\mathcal{U}$ be
\emph{universe of patches}~$\mathcal{U}=\mathcal{M}\cup\mathcal{C}$.
We can define sim as:

\begin{equation}
\operatorname{sim}: \mathcal{U}\times\mathcal{U}\to\interval{0}{1}
\end{equation}

\noindent where 0 denotes no similarity, and 1 denotes textual equivalence.
The operator sim considers various tuneable aspects for the comparison of two patches.
We choose a method due to Ramsauer~et~al.~\cite{ramsauer2019list} that provides high accuracy.

The overall rating for the similarity of two patches in this approach consists of a similarity score of the commit message and a similarity score for the diff.
Both are weighted by a heuristic factor.
The score for the comparison of commit messages and diffs is mainly based on token-based Levenshtein string distances~\cite{levenshtein} (for other
details that are not relevant for our purpose, we refer to Ref.~\cite{ramsauer2019list}).

The approach turns assigning patches in mails to commits in repositories to a problem in graph theory:
The universe~$\mathcal{U}$~forms the vertices of an undirected and weighted graph $G=(\mathcal{U}, E)$.
The weight $w(e)$ of an edge $e=\{e_1,e_2\}\in E$ is determined by a function $\text{sim}$ that rates the similarity of two patches $e_1$ and $e_2$:

\begin{equation}
w(e) := \operatorname{sim}(e_1, e_2)
\end{equation}

Determining all edges in $E$ requires $\binom{|\mathcal{U}|}{2}$ computational expensive calls of sim.
Therefore, prefiltering strategies mitigate the combinatorial explosion for a high number of patches $|\mathcal{U}|$:
Patches are only compared if they, for example, modify at least one common file.
Further prefiltering methods are explained in~\cite{ramsauer2019list}.

The graph $G$ is used to derive an undirected and unweighted subgraph $G'=(\mathcal{U}, E')$ that only contains edges exceeding a certain threshold
$t$ for the edge weight:

\begin{equation}
E'=\{\{e_1,e_2\}\in\mathcal{U}\mid\operatorname{sim}(e_1,e_2)>t\}
\end{equation}

\begin{figure}
    \input{img/graph}
    \caption{Creating clusters of similar patches.
    Patches that exceed the threshold $t=0.8$ form subgraphs of similar patches. Cluster $\alpha$: contains patches on mailing lists as well as commits, cluster $\beta$  only contains patches on mailing lists. Cluster $\gamma$ is not mapped to any mail artefact and a potential off-list patch.}
    \label{fig:graph}
\end{figure}
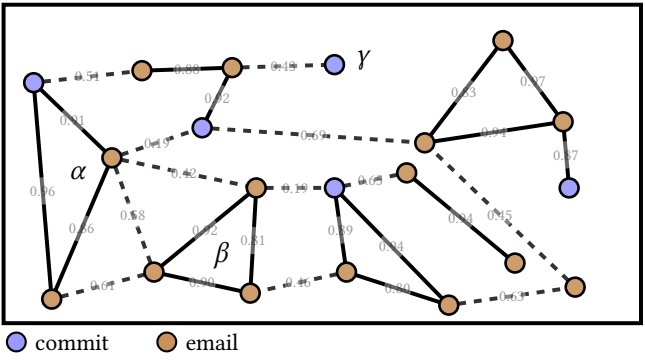

$G'$ consists of connected components that divide $\mathcal{U}$ into partitions of similar patches, that is, equivalence classes.
We identify those equivalence classes as $\sim_S$:
\begin{equation}
[x]_S = \{ y \in \mathcal{U} \mid x \leadsto_{G'} y \},
\end{equation}
where $\leadsto_{G'}$ denotes reachability.
Note that $\left\lvert [x]_S \right\rvert > 0$.
Figure~\ref{fig:graph} illustrates the creation of clusters of similar patches.
We distinguish between three different types of clusters:

\newcounter{eqCtr}\setcounter{eqCtr}{0}
\newcommand\eqclass[2]{
\refstepcounter{eqCtr}
\paragraph{\theeqCtr. #1}$ $\\
#2
}

\eqclass{Unintegrated Patches: $[x]_S\subset\mathcal{M}$}{
All members of this equivalence class can only be found on mailing lists.
Members are, for example, different revisions of the same patch.

No commit in repositories can be found in this category.
This means that the patch has either not been integrated yet (as the discussion is, for example, still ongoing), or, that the patch has been rejected and is not object to integration.
}
\eqclass{Integrated Patches: $\exists x_1, x_2\in[x]_{S} : x_1\in\mathcal{M} \land x_2\in\mathcal{C}$}{
This category describes an finished integration process, as members of the equivalence class can be found on both: the mailing list and the repository.
Again, $[x]_S\cap\mathcal{M}$ describes several revisions of the patch, $[x]_S\cap\mathcal{C}$ denotes the assigned commits in the repository.

Note that several commits in the repository may be assigned:
While backports also match to mainline commit, there may even been multiple mainline commits, as a patch can be picked up by multiple maintainers and appear as multiple commits.
}
\eqclass{Off-list Patches: $[x]_S\subset\mathcal{C}$}{
No public development artefact can be assigned to the commit(s) in $[x]_S$.
Besides false positives results of the heuristic, this category contains commits that arise from non-public integration channels.
}
\\

With patches in group 3, we are able to identify commits that come from secret integration channels.
Those commits will be subject of our analysis in Section~\ref{sec:analysis}.

\subsection{Data Acquisition}
The Linux Kernel community officially provides mailing list archives.\footnote{Available at \url{http://lore.kernel.org}}
As archiving method, they use the public-inbox storage format.\footnote{see \url{https://public-inbox.org/README.html}}
The public-inbox approach stores mails in git repositories and provides a convenient data exchange format as standard tooling can be used to search for or to extract mails from the repository.
Different mailing lists are stored in separate repositories.

However, a low amount of false positives because of misses requires full coverage of all mailing list data for the time frame of interest.
While official resources reach back to early days of Linux, archives do not cover all mailing lists.
Only a subset of $\approx$100 lists of over 200 referenced lists of the project are provided by the Linux Foundation.

Therefore, we subscribed to all 200 publicly available lists and collect mailing list data since May 2019.
Our archives receive regular updates and are publicly available\footnote{Available at \url{https://github.com/orgs/linux-mailinglist-archives/}}.
We use the open-source tool PaStA\footnote{Available at \url{https://github.com/lfd/PaStA}}
for the construction of the commit hash $\leftrightarrow$ message-id map.

%% file: img/graph.tex
\ifdef{\isstandalone}{
	\setlength{\linewidth}{225.84pt}
}

\SetVertexStyle[MinSize = 2.5mm]
\SetPlaneWidth{0.5\linewidth}
\SetPlaneHeight{\linewidth}
\SetPlaneStyle[FillColor=white]
\SetEdgeStyle[TextFillOpacity=0.4, TextFont=\scriptsize, TextFillColor=white]%vertexfill!80]

\begin{tikzpicture}
\Plane[layer=1, opacity=0.8]
\Vertices[NoLabel, layer=1, opacity=0.9]{res/vertices.csv}
\EdgesNotInBG
\Edges[color=black, opacity=1, layer={1,1}]{res/edges.csv}
\Edges[color=black, opacity=0.8, layer={1,1}, style={dashed}]{res/edges_unconnected.csv}

\node at (1, 2) {\Large$\alpha$};
\node at (2.9, 0.9) {\Large$\beta$};
\node at (4.8, 3.5) {\Large$\gamma$};

\node[circle, draw=black, fill=blue!40, inner sep=0pt, minimum size=2.5mm, thick, label=right:{commit}] at (0.17, -0.25) {};
\node[circle, draw=black, fill=brown!85, inner sep=0pt, minimum size=2.5mm, thick, label=right:{email}] at (2.17, -0.25) {};
\end{tikzpicture}

%% file: analysis.tex
\section{Analysis}
\label{sec:analysis}

In contrast to a \emph{just-in-time} online analysis that constantly monitors new incoming mails and commits on a regular basis, we perform the detection of off-list patches as an ex-post analysis of a predefined time window.
From a retrospective view, we can examine if a commit would have been detected as an off-list patch if a just-in-time online analysis would have been performed.

\subsection{Overview}

We are naturally limited by the availability of artefacts for the choice of time window for the analysis.
For the time window of emails, we consider the date since creation of our data collection (2019-May-01) until we performed the analysis (2019-Dec-01).

Patches typically take weeks to months until they are integrated to the repository~\cite{adams13-makeit}.
To select commits in the repository that are relevant for the analysis, we need to be aware that git, the version control system that is used by the Linux kernel, distinguishes between two temporal events:
the author date and the commit date.
The commit date is the date when the commit has been applied to the developer's (local) repository.
Rewriting a repository's history can affect commit dates.
The author date is the date when the commit was originally made (e.g., the date when the code was committed by the original author to their repository) or, in case of an email-based workflow, the timestamp when the email was sent (i.e., the \texttt{Date:} header of a mail).
Hence, we integrate all commits with an author date within the same time window as chosen for emails.
We respect all commits that meet the abovementioned criterion up to Linux version 5.4 (released 2019-Nov-24).

In that time window, we found 516,197 different messages, $\approx40\%$ of them contain actual patches.
However, not all mails that contain patches are relevant for the analysis.
Messages contain mails from bots, pull requests, backports and other noise.
The tool PaStA filters those messages by applying appropriate heuristics.

The remaining messages are first compared against each other to find clusters of similar patches (i.e., several revisions of the same logical patch) and are then compared and mapped against 30,396 commits in the repository.
Thresholds significantly influence the precision of the results.
We chose thresholds in alignment with Ref.~\cite{ramsauer2019list}.
In the time window of our analysis, we mapped $\approx 96\%$ of all commits against patches from mailing list and therefore regular development noise,
while 1,240 commits were not assigned to any Message-Id.

A commit with a missing mapping to a message can fall into one of the following categories:
(1) The heuristic failed to detect the patch (false negative)
(2) The original patch was sent to the list before we started recording mailing list data (miss of discussions)
(3) \emph{off-list patches} -- patches that were integrated through a non-public channel.

\input{evaluation}

%% file: evaluation.tex
\subsection{Off-list Patches}

\begin{figure}
    \includegraphics{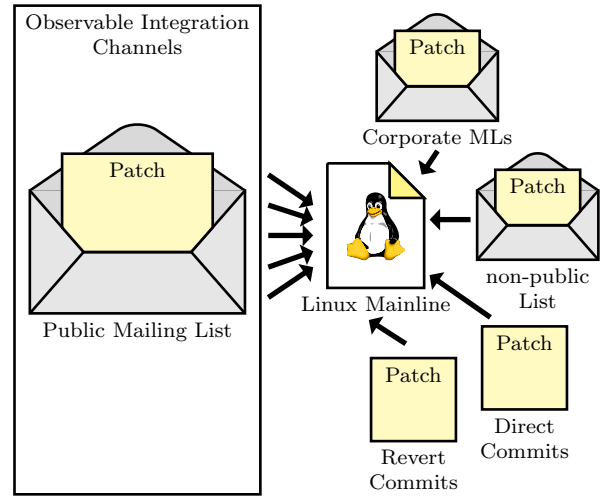}
    \caption{Public observable (left) and non-public integration channels (right).}
    \label{fig:integration}
\end{figure}

With a manual investigation of the remaining 1,240 commits, we were able to find different categories for off-list patches in the Linux kernel repository.
Figure~\ref{fig:integration} illustrates different types of off-list integration channels.

\paragraph{Revert commits}

A revert commit is a commit that reverts a previous commit in the repository's history.
They are used, for instance, to eliminate new features or enhancements if they cause undesired side effects or if they are in a defective or an incomplete state.
It is often the preferred choice to revert the commit, as it is more efficient and less error prone to simply revert corresponding changes rather than to provide expensive or complex fixes, especially at the end of a development cycle.
A refined version of the commit can later be integrated during the next development cycle.\footnote{
Example:
Linux Commit \href{https://git.kernel.org/pub/scm/linux/kernel/git/torvalds/linux.git/commit/?id=69bf4b6b54fb7f52b7ea9ce28d4a360cd5ec956d}{69bf4b6b54fb}:
\emph{[...] and it's [the bug] not immediately obvious why it happens.  It's too late in the rc cycle to do anything but revert for now.}}

Many maintainers do not send reverting patches to mailing lists.
They either integrate the reverting patch directly, or they send a response to the original thread of introducing patch that it will be reverted while omitting the actual reverting patch.
Hence, we lack the reverting patches on mailing lists.

Such reverting patches can automatically be detected, as the subject line of the commit message contains the keyword \texttt{Revert} by convention.
For the time window of the analysis, we detected 64 off-list revert commits in the repository.

\paragraph{Commits by Repository Owners}

Repository owners have a special role in projects:
They have the permission to push code to official resources.

In case of Linux, Linus Torvalds is the owner of the official repository and the last approving authority.
It is his final decision to judge if a patch or pull request is integrated mainline.

This, in turn, allows him for integrating or reverting patches ad libitum.
It is not unusual that he reverts patches without discussion or short before the release of a new version.\footnote{
Example:
\href{https://lkml.org/lkml/2019/9/15/241}{Release of Linux 5.3}:
Linux Commit
\href{https://git.kernel.org/pub/scm/linux/kernel/git/torvalds/linux.git/commit/?id=72dbcf72156641fde4d8ea401e977341bfd35a05}{72dbcf7215}
}

Torvalds sees himself as the manager of the Linux kernel -- and no longer as active developer.
Nevertheless, he sometimes integrates code or fixes without any prior public discussion.
We can automatically detect those commits, as project owners are known.
For the time window of the analysis, we detected 48 off-list patches from Linus Torvalds.
None of them contained security-related fixes.
\\

\noindent However, the phenomenon of bypassing public review processes can also be observed at other maintainers.

\paragraph{Bypass of public review processes}

During our analysis, we found several regular patches that have never been sent to any public mailing list.
To exclude false negatives of the heuristic, we contacted 18 different authors and collected affirmative answers from 14 authors -- four did not answer.

All of them confirmed our finding that their patch(es) have never been posted on a public mailing list.
For example, we found that one maintainer committed 40 patches to their repository in the time window of our analysis.
The author confirmed our assumption and commented that they did not expect it to be that many.
Most patches were only minor stylistic fixes, but we also found invasive patches.
The author agreed that those patches would have required a public review process.
We questioned maintainers why they skipped the official review process.
Their typical answers was that they 
\emph{accidentally forgot to send the patch}.

While many of those commits contain uncontroversial changes like documentation, style or typographical fixes, other commits contain in-depth fixes for subsystems.
One maintainer explained that they picked up a fix from another subsystem that is also valuable for their area of responsibility.
However, all responding maintainers agreed that those patches should have been publicly discussed.

\paragraph{Established non-public integration channels}

Besides maintainers that directly commit patches without discussion, we also found subsystems that tend to bypass public review processes.

We have evidence from our observations that some subsystems deliberately bypass public review processes.
For example, there are whole architectures and subsystems that are in the responsibility of certain companies.\footnote{We do not want publicly point to those subsystems.}
A corporate representative has the role as an official mainline subsystem maintainer, which gives them the possibility to send pull requests to Linus---by trust.
Within those subsystems, we can find off-list patches from authors other than the maintainer.
Still, those patches can not be found on any public mailing lists.
At the same, the author's and maintainer's email address show that both work for the same company.

From such artefacts in commits, we conclude the existence of non-public company internal review and integration processes.
However, those patches do intentionally bypass public review process.

One maintainer confirmed our assumption and underlined that they \emph{forgot to add the public list}, and that \emph{normally all patches are discussed on the public mailing list before they land}.

\paragraph{Security Vulnerabilities}

The remaining commits contain fixes for security vulnerabilities.
According to Linux's security process (explained in Section~\ref{sec:process}), patches for security vulnerabilities should be discussed on private non-public communication channels.

Typically, the majority of those patches drown in the noise of thousands of other commits.
To prevent simple keyword-based search heuristics, commit messages are worded neutrally, links to CVE entries are only sometimes mentioned in the commit message~\cite{greg2019-mds}.

To confirm our assumption that we hit security vulnerabilities through non-public integration channels, we contacted 12 authors.
A list of the related and confirmed vulnerabilities can be found in Table~\ref{tab:vulns}.
All of them confirmed that those patches are security related and that they have either been discussed on the non-public security mailing list, or been sent directly to the maintainer.

We calculated, in days, how long it takes for Debian 10 (Buster) and Ubuntu 18.04 (Bionic Beaver Hardware Enablement Kernel) to apply the patch to the distribution's fork of the Linux kernel.
Positive numbers denote a potential temporal advantage for an attacker, negative numbers mean that the distribution applied the patch before it was disclosed to public.
The categories of vulnerabilities contain denial of service attacks, buffer overflows, privilege escalation, and buffer over-reads.

In our analysis, we found, among others, fixes for the spectre-like attacks CVE-2019-11135~\cite{Schwarz2019ZombieLoad} and CVE-2019-1125~\cite{cve-1125}.
Ubuntu integrated both fixes before they were publicly disclosed, while Debian only integrated fixes for CVE-2019-11135 before they were publicly disclosed.

We also found patches for an \emph{easy to exploit}\footnote{According to an assessment by the author of the fixes~\cite{zyngier19-private}} denial-of-service attack for ARM64-based Cavium systems.
The vulnerability has no assigned CVE entry.
It took almost two months for Ubuntu Bionic to integrate the patch.
At the time of writing, Debian Buster, as well as the affected 4.19 Linux LTS tree, still lack appropriate fixes.

For the majority of vulnerabilities, our approach gives an attacker a temporal advantage from 2 to 179 days.
While most patches for vulnerabilities are included on the stable Linux LTS trees, some distributions still lack patches for the corresponding vulnerabilities.

\input{vulns}

%% file: vulns.tex
\begin{table*}[t]
    \caption{A list of vulnerabilities that were detected by our approach and confirmed by corresponding authors. A negative period means that patches were integrated by distributions before the vulnerabilities were disclosed. SecML means if the patch was routed through the security mailing list, or privately discussed with maintainers.}
    \begin{tabular}{llrlll}
    \toprule
    CVE-2019         & Description & Patches & SecML & Ubuntu 18.04 & Debian 10 \\
    \midrule
    % Zyngier
    NA         & DoS vulnerability for Cavium systems     & 4 & no & 59d & n/a  \\
    
    % Horn
    NA         & smack: use after free & 1 & no           & 14d & n/a \\
    13233 & x86/insn-eval: use after free & 1 & yes  & 54d & 61d \\
    13272 & pot. privilege escalation & 1 & yes      & 25d & 12d \\
    
    % Ellermann
    12817 & ppc: inter-process memory leak & 2 & yes & -5d & 44d \\
    
    % Poimboeuf et al
    1125 & x86/speculation: spectre v1 swapgs & 4 & yes & -5d & 2d \\
    
    % Eremov
    14283 & floppy: out-of-bounds read & 2 & yes        & 71d & 18d \\
    14283 & floppy: DoS / div by zero & 2 & yes         & 71d & 18d \\
    
    % Sriram Rajagopalan
    11833 & ext4: leak of sensitive data & 1 & yes      & 71d  & 29d \\
    
    % ZENG
    NA & s390: pot. leak of sensitive data & 1 & yes         & 45d & n/a \\
    
    % Johansen
    NA & apparmor: out of bounds by user-controlled data & 1 & no & 179d & 60d \\
    
    % Gupta / Hock
    11135 & x86/tsx/speculation: TSX async abort side channel & 9 & yes & -1d & -1d \\
    
    \bottomrule
    \end{tabular}
    \label{tab:vulns}
\end{table*}

%% file: discussion.tex
\section{Discussion}
\label{sec:discussion}
In this section, we first examine validity and potential weaknesses of our approach,
and then discuss how our results affect OSS development processes. We 
conclude with suggestions how they can be adapted to accept (and deal with) risks that are anyway unavoidable, and concentrate on handling highly critical issues as good as possible.

\input{validity}
\input{consequences}

%% file: validity.tex
\subsection{Validity}

\paragraph{Analysis method}
Our work conducts an ex-post analysis.
We consider a time window of seven months worth of mailing list data, and compare it against the corresponding time window in the repository.
This allows us to judge from a \emph{future} perspective if a patch \emph{would have} been detected as an off-list patch at the time it was integrated into the repository.

Nevertheless, the retrospective position is only required to determine the practicability of the approach:
It is straightforward to extended our methods to apply just-in-time, which
is obviously necessary to abuse any undistributed security fixes.
Periodic, frequent updates of the repository and mailing list data ascertain valid and current
data, and are a mere technical detail.
New incoming commits must be compared against the available mailing list data.
If a patch is not an off-list patch, then the corresponding mailing list entry must be available at the moment of the analysis.
As soon as a commit is pushed to a public available repository, our method allows to determine if the commit comes from a private channel.

In a private discussion, Greg Kroah-Hartman, maintainer, among others, of the stable and LTS trees of the Linux kernel, states the undocumented procedures how patches are distributed behind the scenes~\cite{greg2020-private}.
The exchange strategies vary depending on the involved maintainer(s) and the issue at hand:
One possibility is to exchange patches via private email.
Another method is to distribute patches as git bundles, a technique that allows for exchanging elements of a git repository without relying on public remote servers, while it still guarantees stable commit hashes to maintain unique patch identifiers.
In a third method Linus Torvalds pulls patches from a maintainer tree.
Since such trees are publicly available, this method opens a further temporal advantage for attackers, as a just-in-time analysis can also monitor patches from maintainers' repositories.

\paragraph{Generalisability}
The primary concern of this paper is an in-depth analysis of patch flow into the Linux kernel repository from non-public resources by using peculiarities of its mail-based development process.
However, our approach is neither limited to Linux as analysis target, nor to mailing lists as means of discussion.
Variants of this process are used by many other system-level OSS projects, for instance GCC, QEMU, U-Boot, LLVM, busybox, and many others.
Except for handling some technical details and taking minor process differences (e.g., the use of multiple parallel communication channels) into account, our approach can be directly applied to such systems, albeit we do not consider an according evaluation in this paper.

The exact reasons for the existence of non-public integration channels depend on the project.
Especially in projects with smaller communities, maintainers often tend to directly commit code changes without public announcement or discussion (e.g., busybox), as upfront public discussion is often considered time-consuming and dispensable.
However, this limitation is mitigated by the fact that projects with smaller communities only receive a moderate amount of patches.
Especially critical system software typically demands adherence to public review processes, regardless of community size.

Our idea of development process reverse engineering is also applicable to processes that do not build upon mailing lists:
If \emph{any} publicly available development artefacts (e.g., pull requests, entries in issue trackers, \dots) are available that include relevant data before their integration, then reverse process engineering uncovers any irregularities, in particular, deliberate violations of the development process.

\paragraph{Scalability}
In a time window of roughly seven months, we found 30,396 relevant commits in the
repository (authored after 2019-May-01 and integrated before Linux v5.4, released
2019-Nov-24). Within those commits, we found 1,240 potential off-list patches.
By applying heuristics to exclude revert patches and commits by project owners,
we were able to exclude further 112 commits. With our approach, we filtered
$\approx$96\% of regular development noise.

Nevertheless, 1,128 commits required manual analysis, which may
seem to imply a considerable impediment to a fully automatic system at first glance.
However, commits span a time window of 207 days. On a daily basis, this accounts to
manual investigation of (rounded up) six commits per day. Assuming, in  accordance to our personal experience gathered, that an experienced developer can decide within a minute or two if a patch addresses a vulnerability, then the daily time investment would only require a reasonable amount~\cite{Murphy-Hill:2019} of around ten minutes.

% ralf@sd6:linux$ git sl --no-merges v5.0..v5.5 | wc -l
% 70632
% Sun Mar 3 15:21:29 2019 -0800
% Sun Jan 26 16:23:03 2020 -0800
% Delta: 329 days
% -> 214.6 patches / day
Not enjoying the benefits of our system would require a fully manual inspection of all
incoming commits, which is unrealistic: The official repository of the Linux kernel
(merge commits are already excluded) received 70,632 commits between release v5.0 and v5.4.
The development between those releases took 329 days.
On average, 215 patches were integrated per day. Assuming the same amount of time
required for manual investigation, an experienced developer would need more than three
hours of concentrated reviewing per day. Hence, we argue that our approach is suitable
for real-world scenarios, as it significantly reduces the amount of time that is required for manual review.

However, the time to find \emph{some} security-related fixes could be reduced even to zero by employing simple heuristics, such as filtering for well-known author or institution names:
For instance, out of the 12 fixes we identified, 3 originated from Jann Horn (GPZ).
While this might have been pure coincidence, we argue that learning about the social structure behind Linux could be exploited in this respect.

\paragraph{Internal Validity}
Our method systematically uncovers non-public integration channels and identifies
commits that are potential fixes for security vulnerabilities. However, the method fails
for vulnerabilities that are discussed in public before integration.

Statistical data on how many patches are sent to private security mailing lists, or how many critical vulnerabilities are discussed in public are not available.
Hence, it is hard to calculate the accuracy of the approach since the recall is not
available. Yet, we found 12 vulnerabilities in our analysis, which underlines the
practical utility of our approach.

However, it is worth mentioning that counting or searching for CVE entries for a certain time window is neither an appropriate method of accounting the number of vulnerabilities in a system nor an alternative method to automatically find security vulnerabilities:
Only a fraction of kernel security fixes get CVEs~\cite{greg2019-mds, jake2019-cve} assigned.
CVEs are also known to be abused as \emph{integration shortcuts}~\cite{greg2020-private},\footnote{For instances, processes of commercial companies that must be passed before contributions can be placed in open source projects can contain shortcuts for critical vulnerabilities,
and ``critical'' is equated with ``has CVE assigned''.} and do on occasion not even address real vulnerabilities~\cite{greg2019-cve}.

\paragraph{Construct Validity}
We discussed our method with experts of the closed Linux security mailing list.
They confirmed validity of our approach to gain information on non-public integration
channels.

%% file: consequences.tex
\subsection{Consequences}
\paragraph{Fixes for Vulnerabilities}
The primary success criterion for our approach is simple: Can attackers gain temporal
advance to design exploits? We argue that this is the case if the patch can be found in
public resources before software distributors roll out patches: Reverse engineering 
of the development process allows for aimed targeting of commits that would otherwise \emph{hide} between thousands of other commits.

As mentioned in Section~\ref{sec:process}, the majority of patches for vulnerabilities  first appear in the Linux mainline and stable trees before distributions pick up the relevant patches.
From a temporal perspective, patches first appear on mainline and stable trees, and are then integrated by distributions (cf.~\ref{fig:backport}, Vulnerability 1).
We call this the \emph{mainline first} disclosure model.

However, there is an exception for highly critical vulnerabilities:
Before their public disclosure, patches are secretly disclosed to the kernel maintainers
of the distributions, which buys them time to prepare their  kernel tree to roll out
updates (cf. Figure~\ref{fig:backport}, Vulnerability 2) as soon as an embargo ends.
In this way, a patch can be integrated to the distribution's tree before it is
published mainline.

This method ensures that affected systems can receive fixes as soon as the
vulnerability is officially disclosed. Yet, this process requires time-consuming and
extensive coordination between maintainers of distributions and the kernel
community, since a strict temporal publishing coordination is required
to make the approach effective.
Coordination efforts are even more complex when hardware bugs (such as bugs in speculative execution~\cite{Kocher2018spectre, Lipp2018meltdown}) are involved, as multiple operating systems can be affected.
This additionally requires cross-community coordination---between different operating systems (variants of BSD, Windows, macOS), commercial and non-commercial vendors,
and, under exceptional circumstances~\cite{Kocher2018spectre}, even with compiler
manufacturers. This process is therefore only considered in rare cases.

We call this process the \emph{distro first} disclosure model, as patches are integrated by
distributions before they are officially published mainline.

According to Kroah-Hartman~\cite{greg2020-private}, there is no clear definition of the
disclosure process, and no definitive criteria for circumstances when the \emph{distro
first} model should be used. As an ad-hoc process, subsystem maintainers decide how to
handle a fix: patches can, for example, be routed through maintainer trees to Linus
Torvalds, or Linus merges the patch directly, depending on the area of the
kernel that was involved.

To give an example, fixes for flaws in the speculative execution model (cf.\ CVE-2019-11135~\cite{Schwarz2019ZombieLoad} and CVE-2019-1125~\cite{cve-1125}) of modern CPUs were entirely developed and rolled out to distributions in private.
Our approach can still detect that the patches stem from off-list channels as soon as
they are available in a repository -- but at that point in time, patched binaries are
already available for the public. Nevertheless, our method can still provide some valuable
temporal advance as the availability of patches does not imply immediate
deployment in the field.

However, the \emph{mainline first} disclosure model is used for the majority of fixes for
vulnerabilities. As distributions maintain forks of the Linux kernel, and manually select patches
that are integrated from mainline, it can take up to months for patches to be
integrated (cf.~Table~\ref{tab:vulns}). In particular, selecting patches for local forks on
a case-by-case basis misses relevant fixes that are available on LTS.

For these cases, the integration process of distribution kernels can be considered as
\emph{security by obscurity}, since
(a) the patches do not follow a coordinated disclosure process to distributions to
    protect affected systems before their official publication, and
(b) the existence of the actual fixes is obfuscated by private discussion and regular
    development noise.

We hence argue that release strategies of distributions should be reconsidered, as we have demonstrated that distributions are vulnerable for attacks over long periods of time.

Furthermore, we argue that fixes for vulnerabilities should be publicly discussed 
\emph{after} their disclosure.
While preliminary versions for severe vulnerabilities that require \emph{distro first} integration should be developed under the \emph{distro first} model, we recommend
using a full disclosure model in all other cases. Early versions of fixes for vulnerabilities
can still be discussed on secret lists, but they should be publicly reviewed after their embargo.

A public review process can enhance the software quality of the fix per se---after all,
this is the main concern of public discussion---, but can also avoid the the inadvertent introduction  of \emph{additional} vulnerabilities by fixing one
vulnerability, which is unfortunately a real pattern~\cite{greg2019-cve}.
Public discussion before integration would also defeat our mechanisms, which is eventually
desirable.

\paragraph{Code Infiltration}
In addition to detecting fixes for vulnerabilities, we also encountered hidden integration
channels \emph{besides} security mailing lists, such as maintainers or companies that---systematically or inadvertently---bypass official submission procedures, for instance
by direct maintainer commits without external review, or company-internal review.
The existence of such channels, shows that trusted individuals can easily infiltrate the project, and secretly introduce malicious artefacts (while this possibility is given, our method allows for finding concrete instances, which is otherwise not possible).
The existence of such commits contradicts one of the key promises of an open development
model.

We contacted maintainers for subsystems for which we found such patches, and they confirmed our assumption that they integrated code without prior public review.
While maintainers are aware of that they sometimes intentionally bypass the process, they were surprised of the magnitude of unreviewed patches---the confirmed ``record'' is more than 40 per half-year per author, the estimated number for unconfirmed cases is higher.

%% file: related.tex
\section{Related Work}
\label{sec:related}

Software vulnerability life cycle analysis is related to this area work, and a well-researched topic~\cite{huang2016talos, shahzad2012large, arora2010behavior}.

Huang et al.~\cite{huang2016talos} find a considerable delay between disclosure of vulnerabilities and the availability of fixes.
Based on a case study of six different projects, they
found an average time of 52 days from vulnerability disclosure to releasing an actual fix.
However, they also find that almost half of the vulnerabilities are fixed within one week.

In 2010, Arora et al.~\cite{arora2010behavior} argue that instant disclosure of a vulnerability forces vendors to speed up the release of a fix by 35 days.

Shahzhad et al.~\cite{shahzad2012large} analyse the life cycle of vulnerabilities that are filed in software vulnerability data sets.
In their large-scale analysis that includes a big variety of different projects, they find that the amount of time required to fix vulnerabilities decreased from 1998 to 2011:
Since 2008, 80\% of all vulnerabilities are fixed by vendors before their disclosure.
Yet, their study does not consider that providing a patches is only a first step,
but necessitates integration in software distributions, and actual deployment
by users.

In a large-scale empirical study, Li and Paxson~\cite{li2017large} investigate bug-fixes for security vulnerabilities in open-source projects.
For their comprehensive analysis, they assign 3,094 CVE entries in the National Vulnerability Database (NVD) to 4,080 commits in 682 unique git repositories.
Mining for links to commits in the CVE description establishes the approximate connection between CVE entry and commit hash.
Later, they extract characteristics of security-related commits.
They find that security fixes are less complex and more localised than non-security fixes.
Furthermore, they find that 70\% of security-related patches were committed before public disclosure and conclude that development and deployment processes provide a window of opportunity for exploitation.
However, for a responsible disclosure process, it is necessary that patches must be developed (and committed) before disclosure.
Yet, the date of a commit is not necessarily the date of its public visibility.
In this work, we showed that developers intentionally distribute and release patches on secret channels before they finally publicly publish the repositories.
Attackers do not have the opportunity for prior exploitation in those cases.
In this work, we respect this fact and use the time difference of the public availability of a binary software release and the date of the public disclosure as the basis for our analysis.

Kroah-Hartman argues that only a small fraction of Linux kernel security fixes are assigned to CVE entries~\cite{greg2019-mds}.
From 2006-2018, 1005 CVEs were assigned to the kernel.
He argues that, on average, bugs with CVE entries are 100 days fixed in mainline before they get a CVE assigned.
Furthermore, he argues that the amount of vulnerabilities of vendor distributions can significantly be reduced by choosing LTS versions of Linux.

Insider attacks, such as infiltration, or compromises of organisational structures,
are well-known in literature~\cite{bishop2014insider, kammuller2013invalidating}.
We showed a practical \emph{outsider attack} that exploits the openness of the development model itself by using its development artefacts to conclude to systematic integration of
patches that lack public discussion. In~\cite{anderson2002security}, Anderson argues that the
security of a development model should \emph{not} depend whether it is open or closed.

The software engineering community uses artefact mining techniques to to draw quantitative
conclusions on development processes~\cite{joblin2017classifying} or to determine various
software performance indicators~\cite{hemmati2013msr, ferreira2017-refactoring}.

%% file: conclusion.tex
\section{Conclusion}
\label{sec:conclusion}

We showed that reverse engineering of public development processes allows to detect code that arises from non-public integration channels.
Our approach removes 96\% of regular development noise and points to hot spots that contain fixes for critical security vulnerabilities.
With our method, we were able to detect 12 vulnerabilities in Linux in a time window of seven months.
We collected responses from all authors that confirm our presumptions.
Attackers can use this information to gain temporal advantage, as they can design exploits before affected systems receive patches.

Furthermore, we found evidence that some subsystems and maintainers intentionally bypass the regular development process.
Therefore we argue that it is possible to systematically infiltrate malicious code to the kernel by bypassing the (mandatory) public review processes.
We shared our findings with the Linux kernel community and discussed possibilities of potential mitigations.

Our future work will focus on just-in-time online analyses and automated process monitoring:
Automatic notifications to maintainers or authors can help to raise the awareness of the importance of public code review processes.

%% file: ack.tex
\section{Acknowledgements}
We thank Greg Kroah-Hartman for giving us the opportunity to discuss the topic with Linux kernel security officers.
We also thank authors of off-list patches for their detailed answers, discussions on their patches and fruitful conversation.
We do not mention them by name.

We also thank Christian Weber and Florian Loher for their assistance with creating figures and illustrations.

This work was supported by the iDev40 project and the German Research Council (DFG) under grant no. LO 1719/3-1.
The iDev40 project has received funding from the ECSEL Joint Undertaking (JU) under grant no.~783163.
The JU receives support from the European Union's Horizon~2020 research and innovation programme.
It is co-funded by the consortium members, grants from Austria, Germany, Belgium, Italy, Spain and Romania.